\documentclass[a4paper,12pt]{article}
\usepackage{epsfig,a4wide,amsmath,amssymb,hyperref,graphicx}
\usepackage[hang,nooneline]{subfigure}
\title{\bf \Huge Boson Stars in a Theory of Complex Scalar Fields coupled to $U(1)$ Gauge Field and Gravity} 	
\author{{\bf  Sanjeev Kumar$^{a}$\footnote{Corresponding Author} , Usha Kulshreshtha$^b$,}
\\{\bf and Daya Shankar Kulshreshtha$^a$}
\footnote{Email Addresses: skumar1@physics.du.ac.in (Sanjeev Kumar), ushakulsh@gmail.com (U. Kulshreshtha), dskulsh@gmail.com (D. S. Kulshreshtha)}\\[10pt]
$^a$ Department of Physics and Astrophysics,\\ University of Delhi, Delhi-110007, India\\
$^b$ Department of Physics, Kirori Mal College, \\University of Delhi, Delhi-110007, India}
\date{\today}
\newcounter{saveeqn}
\newcommand{\alpheqn}
{ \setcounter{saveeqn}{\value{equation}}\stepcounter{saveeqn}
 \setcounter{equation}{0}
 \renewcommand{\theequation}{\mbox{\arabic{saveeqn}\alph{equation}}}}
\newcommand{\reseteqn}
{ \setcounter{equation}{\value{saveeqn}}
 \renewcommand{\theequation}{\arabic{equation}}}
\begin{document}
\maketitle
\begin{abstract}
We study boson shells and boson stars in a theory of complex scalar field coupled to the $U(1)$ gauge field $A_{\mu}$ and Einstein gravity  
with the potential: $V(|\Phi|) := \frac{1}{2} m^{2} \left(|\Phi|+ a \right)^2$. This could be considered either as a theory of massive complex scalar field coupled to electromagnetic field and gravity in a conical potential or as a theory in the presence of a potential which is an overlap of a parabolic and a conical potential.
Our theory has a positive cosmological constant $(\Lambda := 4 \pi G m^2 a^2)$. Boson stars are found to come in two types, having either ball-like or shell-like charge density.  We have studied the properties of these solutions and have also determined their domains of existence for some specific values of the parameters of the theory. Similar solutions have also been obtained by Kleihaus, Kunz, Laemmerzahl and List, in a V-shaped scalar potential. 
\end{abstract}
\vspace{2cm}
\newpage
Boson stars and boson shells are hypothetical astronomical objects consisting 
of bosons. They could possibly be detected by gravitational radiation emitted 
e.g., by a pair of co-orbiting boson stars  \cite{orbital,gravity}. They could have 
been possibly formed through gravitational collapse during primordial stages 
of big bang \cite{Madsen}. They have also been proposed as dark matter 
candidates \cite{darkmatter}. Further, just like the super-massive black holes 
for example,  they could even exist in the center of galaxies \cite{super1} and 
they could possibly explain many observed properties of the active galactic 
core \cite{super2}.

The study of boson shells and boson stars in scalar electrodynamics with a self-interacting complex scalar field $\Phi$ coupled  to Einstein gravity in three-space one-time dimensions is of a very wide interest \cite{orbital}-\cite{9},\cite{17}-\cite{20}. Kleihaus, Kunz, Laemmerzahl and List \cite{1,2} have recently studied boson shells harboring black holes and boson stars in scalar electrodynamics coupled to Einstein gravity in three-space one-time dimensions in a V-shaped scalar potential: $V(\Phi \Phi^* )\equiv V(|\Phi|) = \lambda |\Phi|$ (where $\lambda$ is a constant) \cite{1,2}. They have found that the boson stars come in two types, having either ball-like or shell-like charge density \cite{1,2}. They have studied the properties of these solutions and have also determined their domains of existence \cite{1,2}.

In the present work, we study boson shells and boson stars in a theory of complex scalar field coupled to the $U(1)$ gauge field $A_{\mu}$ and Einstein gravity with the potential $V(|\Phi|)$ defined by: 
\begin{equation}
V(|\Phi|) := \frac{1}{2} m^{2} \left(|\Phi|+ a \right)^2\label{potential}
\end{equation}
for $ a > 0$. Here $m$ and $a$ are constant parameters. 
This could be considered either as a theory of massive complex scalar field coupled to electromagnetic field and gravity in a conical potential or as a theory in the presence of a potential which is an overlap of a parabolic and a conical potential.
As $\Phi\rightarrow0$, $V(|\Phi|)\rightarrow \frac{1}{2}m^2 a^2\,(\,:=\Lambda/(8\pi G)\,) $ which can be interpreted as a presence of a positive cosmological constant $\Lambda=4 \pi G m^2 a^2$ in the theory.


We study the properties of boson star and boson shell solutions (with the interior of the shells as an empty space (albeit, boson-shells with the de Sitter like interior)). The numerical procedure used by us called as the shooting method is described, in brief, at the end. 

The action of the theory under consideration reads:
\begin{equation}
S=\int \left[ \frac{R}{16\pi G}
   - \frac{1}{4} F^{\mu\nu} F_{\mu\nu}
   -  \left( D_\mu \Phi \right)^* \left( D^\mu \Phi \right)
 - V( \Phi \Phi^* ) 
 \right] \sqrt{-g} ~~ d^4 \tilde{\tilde x}
  \label{action}
\end{equation}
Here $R$ is the Ricci curvature scalar, $G$ is Newton's Gravitational constant and $e$ is the gauge coupling constant. $D_\mu \Phi = (\partial_\mu \Phi + i e A_\mu \Phi)$ and the electromagnetic field strength tensor is defined as: $F_{\mu\nu} = (\partial_\mu A_\nu - \partial_\nu A_\mu)$. Also, $g = det(g_{\mu\nu})$ where the metric tensor $g_{\mu\nu}$ is defined somewhat later, and the asterisk in the above equations denotes complex conjugation.


 Also, to construct static spherically symmetric solutions we adopt the spherically symmetric metric with Schwarzschild-like coordinates
\begin{eqnarray}
 ds^2&=& g_{\mu\nu} dx^\mu dx^\nu = \biggl[ -A^2(r) N(r) dt^2 + N^{-1}(r) dr^2 +r^2(d\theta^2 + \sin^2 \theta \;d\phi^2) \biggr] \notag \\
{\bf g_{\mu\nu}}&=&diag\,\left(-A^2(r) N(r) ,\; N^{-1}(r) ,\; r^2,\; r^2 \sin^2 \theta\right) 
\end{eqnarray}
The equations of Motion for the fields are obtained by variation of the action with respect to the metric and the matter fields
\begin{eqnarray}
 G_{\mu\nu}&\equiv& R_{\mu\nu}-\frac{1}{2}g_{\mu\nu}R = 8\pi G T_{\mu\nu}
\notag \\ 
 \partial_\mu \left ( \sqrt{-g} F^{\mu\nu} \right)& =&
   -i\, e \sqrt{-g}\, [\Phi^* (D^\nu \Phi)-\Phi (D^\nu \Phi)^* ] 
\notag \\
D_\mu\left(\sqrt{-g}  D^\mu \Phi \right) &=&
    \frac{1}{2}\sqrt{-g}m^2\Phi \left(1+\frac{a}{|\Phi|}\right)
  \label{vfeqH}
 \end{eqnarray}
The equation of motion for the field $\Phi^{*}$ is obtained by the complex conjugation of the last equation. The stress-energy tensor $T_{\mu\nu}$ is given by ,
\begin{eqnarray}
T_{\mu\nu} 
&=& \biggl[ ( F_{\mu\alpha} F_{\nu\beta}\ g^{\alpha\beta}
   -\frac{1}{4} g_{\mu\nu} F_{\alpha\beta} F^{\alpha\beta})
- \frac{1}{2} g_{\mu\nu} \left(     (D_\alpha \Phi)^* (D_\beta \Phi)
  + (D_\beta \Phi)^* (D_\alpha \Phi)    \right) g^{\alpha\beta}
\notag \\
& &  ~~~~ ~~~~ ~~~~ ~~~~ ~~~~ ~~~~ ~~~~ ~~~~
 + (D_\mu \Phi)^* (D_\nu \Phi) + (D_\nu \Phi)^* (D_\mu \Phi)
 -  g_{\mu\nu}\; V( |\Phi|) \biggr] ~~~
  \label{vtmunu}
\end{eqnarray}
We now obtain
\begin{eqnarray}
G_t^t &=& \biggl[ -\frac{1}{r^2}\left[r\left(1-N\right)\right]' \biggr] ~~,~~
G_r^r = \biggl[ \frac{2 r A' N -A\left[r\left(1-N\right)\right]'}{A\ r^2} \biggr] \notag \\
G_\theta^\theta &=& \biggl[ \frac{2r\left[rA'\ N\right]' + \left[A\ r^2 N'\right]'}{2 A\ r^2} \biggr]
\  \   = \  G_\phi^\phi
\end{eqnarray}
Here the arguments of the functions $A(r)$ and $N(r)$ have been suppressed. For solutions with vanishing magnetic field, the Ansatze for the matter fields have the form:
\begin{eqnarray}
 \Phi(x^\mu)=\phi(r) e^{i\omega t}
~~,~~
A_\mu(x^\mu) dx^\mu = A_0(r) dt
\end{eqnarray}
With these Ansatze the Einstein equations
\begin{eqnarray}
\;G_t^t = 8 \pi G\ T_t^t ~~,~~ \;G_r^r =  8 \pi G\  T_r^r ~~,~~ \;G_\theta^\theta =  8 \pi G\ T_\theta^\theta
\end{eqnarray}
where the arguments of $A(r)$, $N(r)$, $\phi(r)$ and 
$A_0(r)$ have been suppressed, reduce to:
\begin{eqnarray}
 \frac{-1}{r^2}\left[r\left(1-N\right)\right]' 
& = & \frac{-8\pi G}{2A^2 N e^2} \left[ N [(\omega +e A_0 )']^2 +(\omega + e A_0 )^2 (\sqrt{2} e \phi)^2\right. \notag \\
& &\hspace{0.1in}\left.+ A^2 N^2 (\sqrt{2} e \phi')^2 + \frac{1}{2}A^2 N m^2(\sqrt{2}\, e\phi +\sqrt{2}\,e\, a)^2 \right] 
\end{eqnarray}
\begin{eqnarray}
\frac{2 r A' N -A\left[r\left(1-N\right)\right]'}{A r^2}
& = &\frac{8\pi G}{2A^2 N e^2} \left[ -N [(\omega +e A_0 )']^2 +(\omega + e A_0 )^2 (\sqrt{2} e \phi)^2\right. \notag \\
& &\hspace{0.5cm}\left.+ A^2 N^2 (\sqrt{2} e \phi')^2 - \frac{1}{2}A^2 N m^2(\sqrt{2}\, e\phi +\sqrt{2}\,e\, a)^2 \right] 
\end{eqnarray}
\begin{eqnarray}
\frac{2r\left[rA'N\right]' + \left[A r^2 N'\right]'}{2 A r^2} & = &\frac{8\pi G}{2A^2 N e^2} \left[ N [(\omega +e A_0 )']^2 +(\omega + e A_0 )^2 (\sqrt{2} e \phi)^2\right. \notag \\
& &\hspace{0.5cm}\left. -A^2 N^2 (\sqrt{2} e \phi')^2 -\frac{1}{2}A^2 N m^2(\sqrt{2}\, e\phi +\sqrt{2}\,e\, a)^2 \right]  \label{dtheta}
\end{eqnarray}
\noindent Here the prime denotes differentiation with respect to $r$ and the equation $G_\phi^\phi =8\pi G\ T_\phi^\phi$ also leads to an equation identical with that of the last equation (Eq. \ref{dtheta}).

For notational simplicity, we introduce new coupling constants and redefine the matter field functions $\phi(r)$ and $A_0(r)$ as follows:
\begin{equation}
 \alpha^2 = \frac{4\pi G\,m^2}{e^2} ~~~,~~~ \tilde{a}=\sqrt{2}\,e\,a
\end{equation}
\begin{equation}
 h(r)=\sqrt{2} \;e \phi(r) ~~~,~~~ b(r)=\omega+eA_0(r).
 \end{equation}
The equations of motion in terms of $h(r)$ and $b(r)$ 
read:
\begin{equation}
\left( A N r^2 h'\right)' = \frac{r^2}{2 A N}\left[A^2 N m^2( h+\tilde{a})\,{\rm sign}(h) -2 b^2 h \right]\label{vheq}
\end{equation}
and 
\begin{equation}
\left[\frac{ r^2\  b'}{A} \right]' = \biggl[ \frac{r^2 h^2 b }{A\ N} \biggr]
\label{vbeq}
\end{equation}
where 
$$ {\rm sign}(h) = \left\{ \begin{array}{ll}
\pm1 \ \ & h>0,~h<0 \\ 0 & h=0 \end{array}\right. $$

Now onwards, $m$ can be removed by the rescaling:
\begin{eqnarray}
h(r)\rightarrow m\, h(m\,r)\,,\;\;\;
 b(r)\rightarrow m\,b(m\,r) \,,\;\;\;\tilde{a}\rightarrow m\,\tilde{a}\, ,\ \ \ \
r\rightarrow m~r \ \ \ \ \Lambda\rightarrow m^2 \Lambda
\end{eqnarray}

With the above Ansatze (with the primes denoting the differentiation with respect to $r$) we obtain:
\alpheqn
\begin{eqnarray}
h''&=& \biggl[ \frac{A^2 N ( h+\tilde{a})\, {\rm sign}(h) -2 b^2 h}{2 A^2 N^2}-\frac{2\,h'}{r}-h'\left(\frac{A'}{A}+\frac{N'}{N}\right) \biggr] ~~~\label{vheq1}  \\
b''&=&\biggl[ \frac{b\,h^2}{N}+\frac{b'A'}{A}-\frac{2\,b'}{r} \biggr] 
\label{vbeq1} \\
\frac{1}{r^2}\left[r\left(1-N\right)\right]'&=& \frac{\alpha^2}{A^2 N}\left(A^2 N^2 h'^2 + N b'^2 +\frac{1}{2} A^2 N(h+\tilde{a})^2+ b^2 h^2\right)
\label{vE_00}\\
\mbox{\hspace{-1.5cm}}
\frac{2 r A' N -A\left[r\left(1-N\right)\right]'}{A r^2} 
&=& \frac{\alpha^2}{A^2 N}
\left(A^2 N^2 h'^2-N b'^2 -\frac{1}{2} A^2 N(h+\tilde{a})^2+ b^2 h^2\right)
\label{vE_rr}\\
\frac{2r\left[rA'N\right]' + \left[A r^2 N'\right]'}{2 A r^2}&=& \frac{\alpha^2}{A^2 N}
\left(-A^2 N^2 h'^2 + N b'^2 -\frac{1}{2} A^2 N (h+\tilde{a})^2+ b^2 h^2\right)
\label{vE_tt}
\end{eqnarray}
\reseteqn

Solving Eqs.(\ref{vE_00}) and (\ref{vE_rr}) for $A'$ and $N'$ and also using eqs (\ref{vheq1}) and (\ref{vbeq1}) we get:
{
\alpheqn
\begin{eqnarray}
N' & = & \frac{1-N}{r} 
-\frac{\alpha^2 r}{A^2 N}
\left(A^2 N^2 h'^2 + N b'^2 +\frac{1}{2} A^2 N (h+\tilde{a})^2+ b^2 h^2\right) \ 
\label{veq_N}\\
A' & = & 
\frac{\alpha^2 r}{A N^2}\left(A^2 N^2 h'^2 + b^2 h^2\right) \ 
\label{veq_A}\\
h'' & = & 
\frac{\alpha^2\, r h'}{A^2N} \left(A^2\,(h+\tilde{a})^2 + \,b'^2\right)
-\frac{h'\left(N+1\right)}{rN}
+\frac{A^2 N ( h+\tilde{a})\, {\rm sign}(h) -2 b^2 h}{2\,A^2 N^2}\hspace{0.2cm}
\label{veq_H}\\
b'' & = & 
\frac{\alpha^2}{A^2 N^2} rb'\left(A^2 N^2 h'^2 + b^2 h^2\right)
-\frac{2 b'}{r} + \frac{b h^2}{N}
\label{veq_b}
\end{eqnarray}
\reseteqn
}
To solve equations (\ref{veq_N}), (\ref{veq_A}), (\ref{veq_H}), (\ref{veq_b}) numerically, we introduce a new coordinate $x$ as follows:  
\begin{equation}
r= r_{\rm i} +x (r_{\rm o}-r_{\rm i})\ , \ \ \ \ 0\leq x \leq 1 \ .
\label{vcoord_x}
\end{equation}
implying that $ r=r_i \ \ at \ \ x=0 $ and 
$ r=r_{\rm o}\ \ at \ \ x=1 $.
Thus the inner and outer boundaries of the shell are always at $x=0$ and $x=1$ respectively, while their radii $r_{\rm i}$ and $r_{\rm o}$ become free parameters.

For the metric function A(r), we choose the 
boundary conditions:
\begin{eqnarray}
A(r_{\rm o})=1 ~~,~~
\end{eqnarray}
where $r_{\rm o}$ is the outer radius of the shell. Since it retains its value upto infinity, this fixes the time coordinate. For constructing globally regular ball like boson star solutions with finite energy, we choose
\begin{eqnarray}
 N(0)=1 ~~,~~
b'(0)=0 ~~,~~  h'(0)=0  ~~,~~ 
h(r_{\rm o})=0 ~~,~~ 
h'(r_{\rm o})=0  \label{bcstar}
\end{eqnarray}
For globally regular boson shell solutions with empty space-time in the interior of the shells, we choose the boundary conditions:
\begin{eqnarray}
N(r_i)=1-\frac{\Lambda}{3} r_i^2 ~,~~
b'(r_{\rm i})=0 ~,~~ h(r_{\rm i})=0 ~,~~h'(r_{\rm i})= 0 ~,~~  h(r_{\rm o}) = 0 ~,~~h'(r_{\rm o})=0\label{bcshell}
\end{eqnarray}
where $r_{\rm i}$ and $r_{\rm o}$ are the inner and outer radii of the shell and $\Lambda = \frac{\alpha^2 \tilde a^2}{2}$ .

The conserved current for boson shell is:
\begin{eqnarray}
j^\mu=-i \,e\,\left\{ \Phi(D^\mu \Phi)^*-\Phi^* (D^\mu \Phi) \right\}\ ~~,~~ 
D_{\mu} j^{\mu} = 0
\end{eqnarray}
The charge contribution of the boson shell to the global charge $Q$ is given by:
\begin{eqnarray}
Q_{\rm sh}=-\frac{1}{4\pi}\int_{r_i} ^{r_0} j^0 \sqrt{-g} \,dr\,d\theta\,d\phi  ~~,~~
j^0=-\frac{h^2(r) b(r)}{A^2(r) N(r)}
\end{eqnarray}
The global charge therefore consists of the charge carried by the bosons forming the shells and the charge localized in the interior $~Q_i~$ of the shells.

The mass $M$ of all gravitating solutions can be obtained from the asymptotic form of their metric. In the units employed, we find
 \begin{equation}
\alpha^2 M= \biggl(1-N(r_o)+\frac{\alpha^2 Q^2}{r_o^2} -\frac{\Lambda}{3} r_o^2\biggr)\frac{r_o}{2}
 \end{equation}

\begin{figure}[!h]
\begin{center}
\mbox{\hspace{-0.5cm}
\subfigure[][]{
\includegraphics[height=.26\textheight, angle =0]{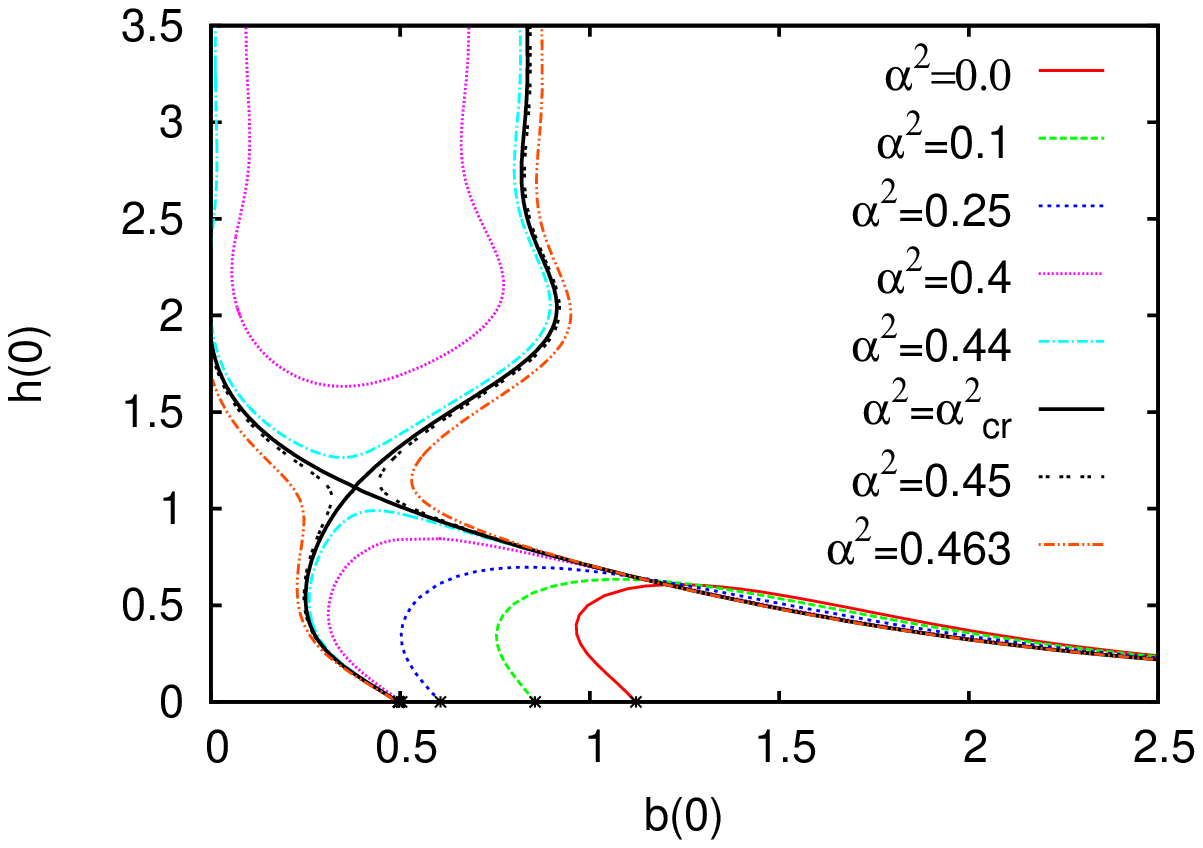}
\label{dphasediag}
}
\subfigure[][]{\hspace{-0.5cm}
\includegraphics[height=.26\textheight, angle =0]{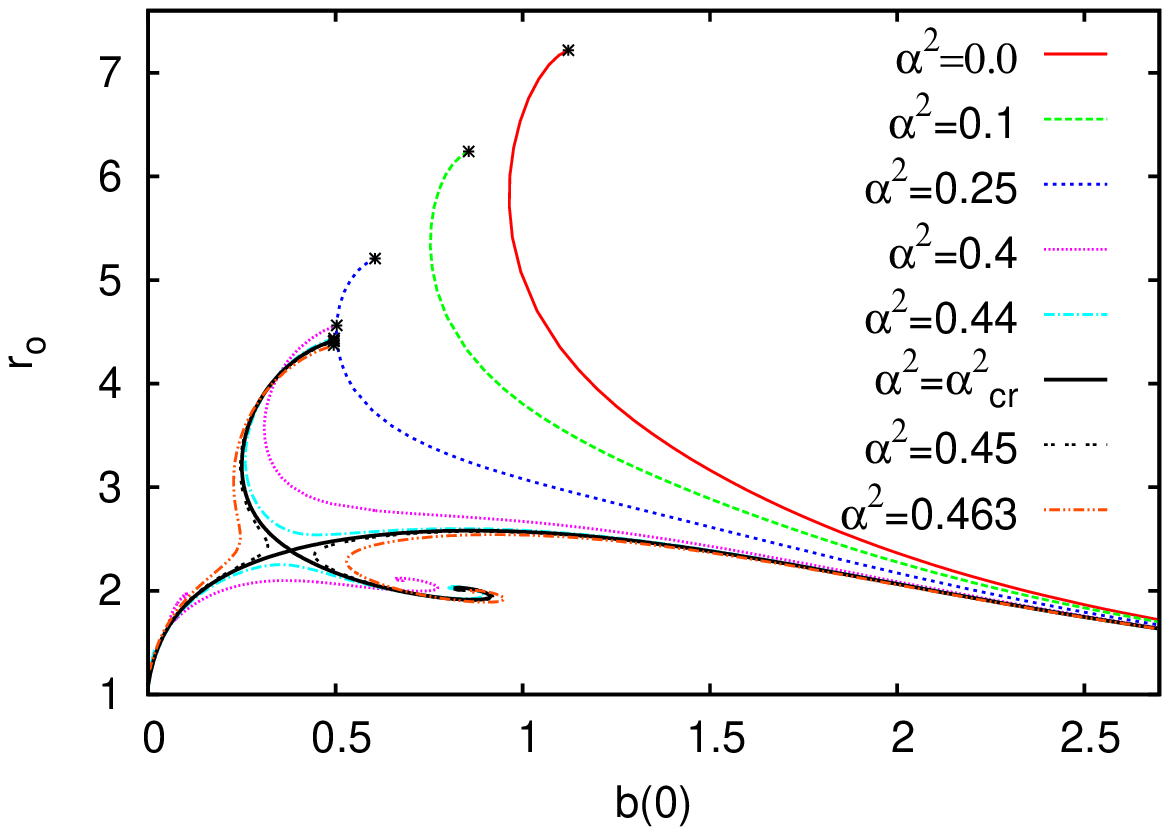}
\label{dr0_vs_b0}
}
}
\vspace{-0.5cm}
\mbox{\hspace{-0.5cm}
\subfigure[][]{
\includegraphics[height=.26\textheight, angle =0]{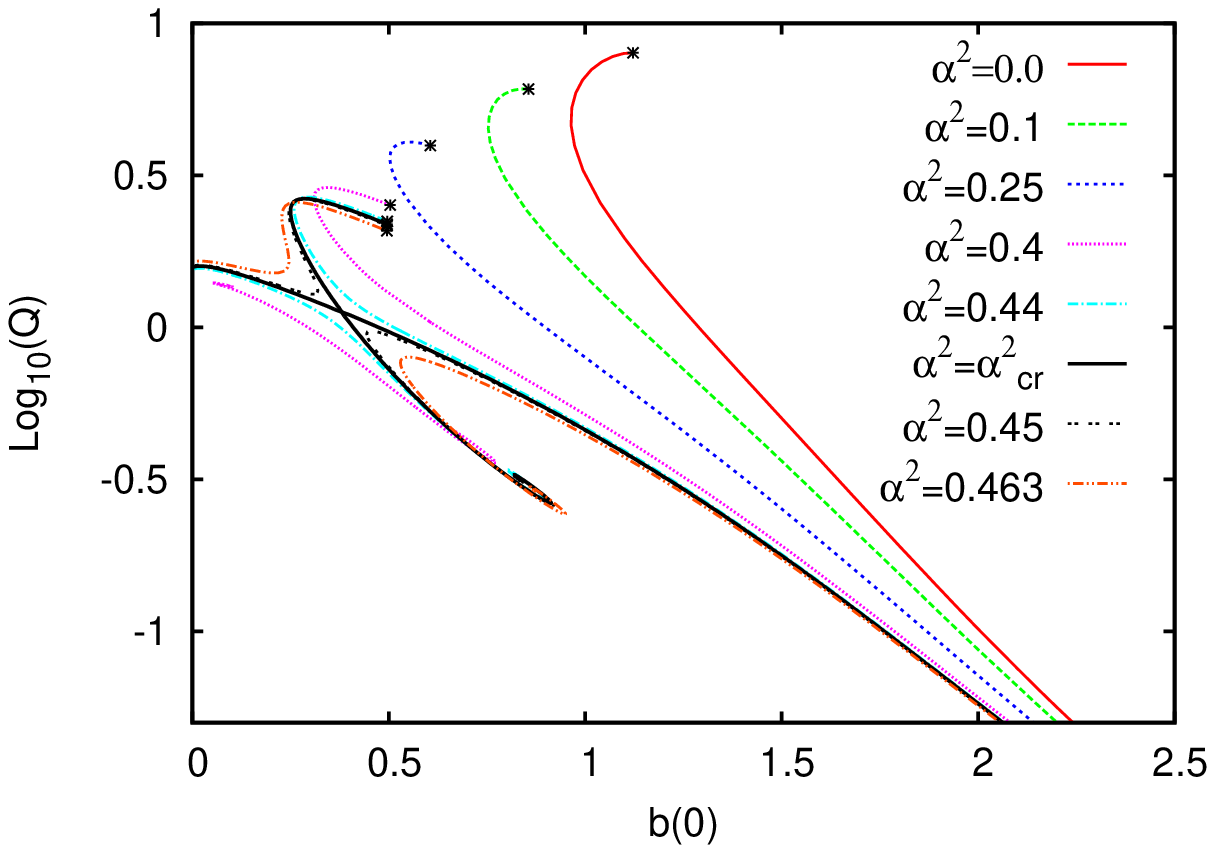}
\label{dQ_vs_b0}
}
\subfigure[][]{\hspace{-0.5cm}
\includegraphics[height=.26\textheight, angle =0]{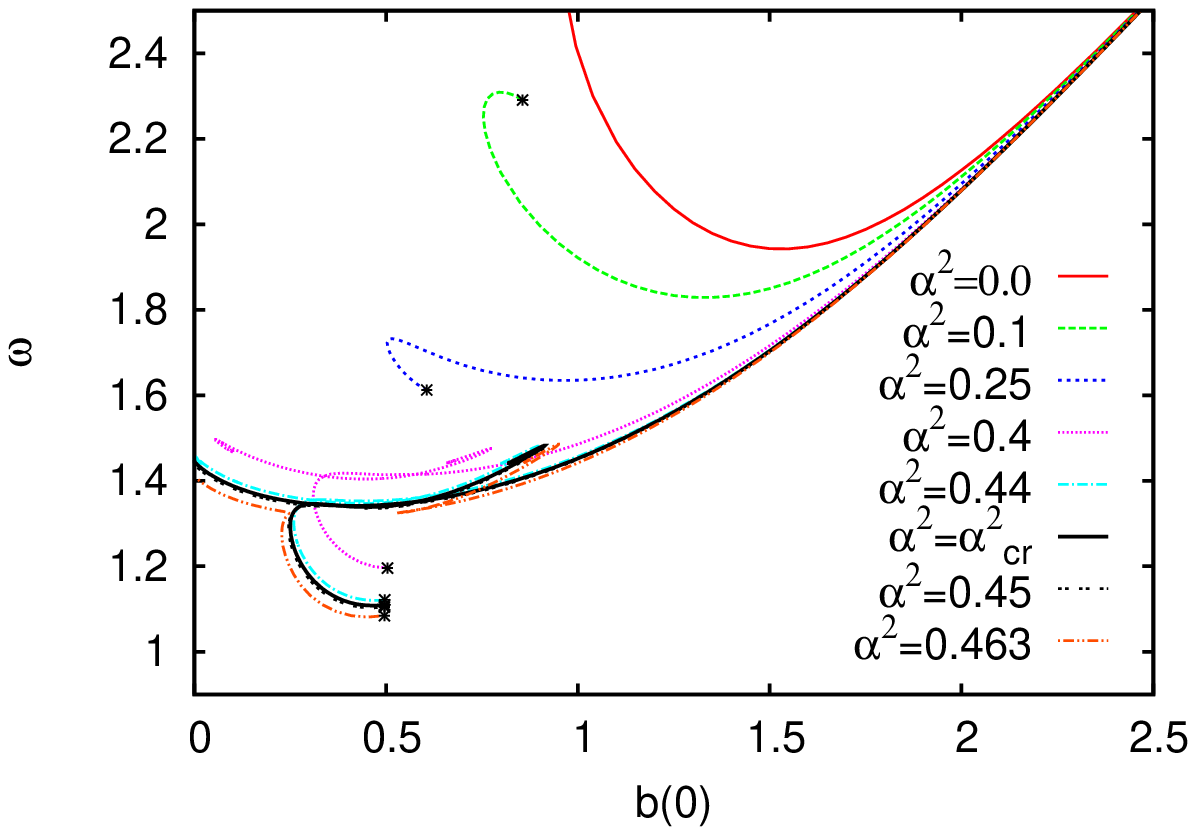}
\label{dw_vs_b0}
}
}
\end{center}
\caption{Properties of the boson star solutions shown versus $b(0)$,
(a) $h(0)$, the value of the scalar field function $h(r)$ at the origin,
(b) $r_o$ indicates the size of the solutions,
(c) exhibit logarithmically the charge $Q$,
and (d) shows the value $b(\infty)$ (which corresponds to the value
of the scalar field frequency $\omega$). The asterisks mark the transition points from boson stars to boson Shells.
\label{dbosonstar}
}
\end{figure}

In the following, we first consider the case of boson stars. 
In Fig.\ref{dphasediag} the sets of boson star solutions for a sequence of values of coupling constant $\alpha$ are given in terms of the values of the scalar field function $h(0)$ and the value of the gauge field function $b(0)$ at the center of the solutions 
(i.e. at $r=r_i = 0$) for a given value of parameter $\tilde{a}=0.5$.

For $\alpha = 0$, the Q-ball solutions form a continuous set, represented by a single curve for smaller values of $h(0)$ and larger values of $b(0)$. These non-gravitating solutions are bounded by some maximal value of $h(0)$, by some minimal value of $b(0)$, and by bifurcation point with the shell-like solutions, where $h(0)$ reaches zero.

As the coupling constant $\alpha$ is increased from zero, the maximal value of $h(0)$ of these sets of solutions then increases, the corresponding minimal value of $b(0)$ decreases, and the bifurcation point with the shell-like solutions decreases as well until a critical value of coupling constant  $\alpha=\alpha_{cr}\simeq0.447$ is reached. Here a second set of solutions is obtained for larger values of $h(0)$ and smaller values of $b(0)$. This second set of solutions if present for each value of $\alpha\le\alpha_{cr}$. For these solutions $h(0)$ has minimal value for fixed $\alpha$, which decreases with increasing $\alpha$, until at $\alpha_{cr}$ the two sets of solutions touch and bifurcate into to other sets of solutions on the left and right of the bifurcation point. In the right region the solutions corresponds to the larger values of $b(0)$, while in the left region they are restricted to the smaller values of the $b(0)$. In the left region the bifurcation point with the shell-like solutions 
decreases with the increase of coupling constant $\alpha$ until a critical value of the coupling constant $\alpha=\alpha_{sh}\simeq0.463$ is reached where the value of $b(0)$ at bifurcation point with the shell-like solutions starts increasing.

Fig.\ref{dr0_vs_b0} shows the outer radius $r_o$ for these sets of solutions, and thus the size of the corresponding boson stars. The oscillations of the gauge field value b(0) with increasing scalar field value h(0) seen in Fig.\ref{dphasediag} are characteristic spirals exhibited by the boson stars. These spirals are reflected in the spirals formed by the outer radius $r_o$ seen in  Fig.\ref{dr0_vs_b0}.
The space-time for $r\geq r_o$ then corresponds to exterior space-time for Reissner Nordstr\"om de Sitter black hole and the metric function N(r) can be expressed as 

\begin{equation}
N(r )= \biggl[ 1-\frac{2 \alpha^2 M}{r} + \frac{\alpha^2 Q^2}{r^2}  -\frac{\alpha^2\tilde a^2}{6} r^2\biggr] 
\end{equation}

The charge $Q$ for these boson star solutions is shown in Fig.\ref{dQ_vs_b0}. The oscillations of b(0) leads to spiral patterns for the charge. The value of the scalar field frequency $\omega$ (which corresponds to $b(\infty)$) is exhibited in figure \ref{dw_vs_b0} and it also remains spiral in form. The Figures \ref{dphasediag}-\ref{dw_vs_b0}) represents the domain of existance of the compact boson star solutions and physical solutions with dimensionfull quantities can be obtained from these dimensionless solutions by appropriate scaling.

\begin{figure}[!h]
\begin{center}
\vspace{-0.5cm}
\includegraphics[width=0.65\textwidth, angle =0]{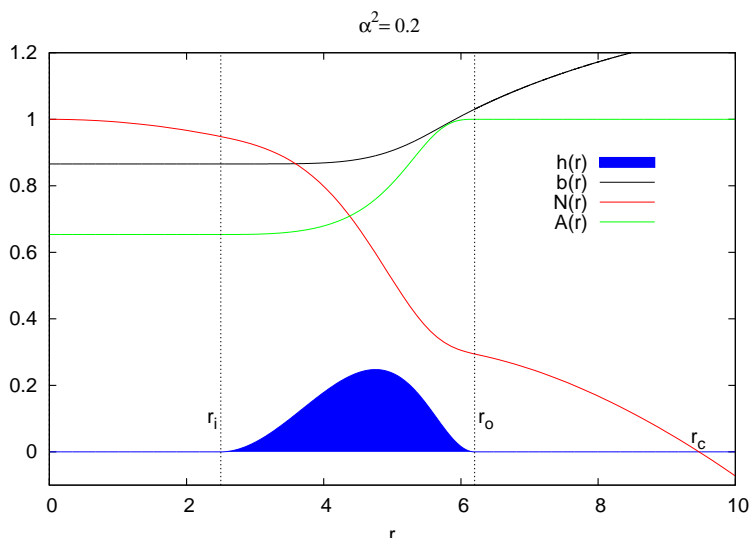}
\vspace{-1cm}
\end{center}
\caption{The field functions h(r), b(r) and metric functions A(r), N(r) are shown for boson shell solutions for $\alpha^2=0.2$ and $\tilde a =0.5$. $r_c$ is the cosmological horizon radius.  }\label{solution}
\end{figure}

For considering the boson shells, we need to distinguish 3 regions of the space-time. In the inner region $0\le r < r_i$ the gauge potential is constant and the scalar field vanishes (cf. Eqs. \ref{bcstar} and \ref{bcshell}). Consequently, it is de Sitter-like, with $N(r) = 1-\frac{\Lambda}{3} r^2$ and $A(r) = const < 1$. The middle region $r_i < r < r_o$ represents the shell of charged boson matter. The outer region $r_o < r < \infty$, corresponds to the outer part of a Reissner-Nordstr\"om de Sitter space-time.
In this outer region the gauge field exhibits the standard Coulomb fall-off, while the scalar field vanishes identically. An example of a solution is shown in Fig. \ref{solution} for $\alpha^2= 0.2$ and $\tilde a =0.5$. with $r_i\,,r_o$ being the inner and outer radii of the shell respectively and cosmological horizon radius $r_c$ corresponds to N(r)=0 for $r=r_c$ .

The domain of existence of these gravitating boson shells depends on the strength of the coupling $\alpha^2$. For a given finite value of the gravitational coupling, boson shells emerge from the boson star solutions, when the scalar field vanishes at the origin: $h(0) = 0$.  The value of the outer radius $r_o$ at the bifurcation point depends on the strength of the coupling $\alpha^2$ for a given value of parameter $\tilde{a}=0.5$.
For the boson shells with de Sitter like interior, when we increase the value of the inner shell radius $r_i$ from zero, while keeping the coupling constant fixed, the corresponding branches of the boson shells are obtained.
 \begin{figure}[!h]
\begin{center}
\mbox{\hspace{-0.2cm}
\subfigure[][]{
\includegraphics[width=0.5\textwidth,height=.27\textheight, angle =0]{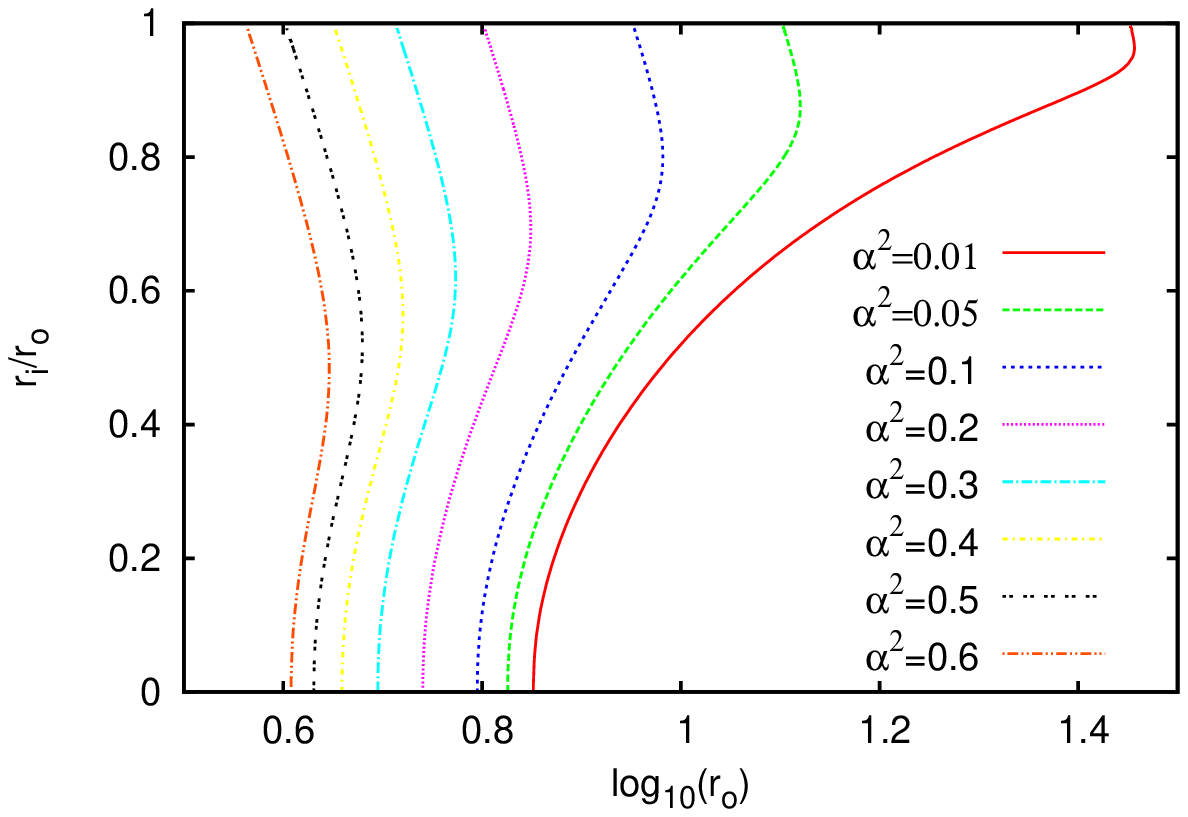}
\label{dSha}
}
\subfigure[][]{\hspace{-0.4cm}
\includegraphics[width=0.5\textwidth,height=.27\textheight, angle =0]{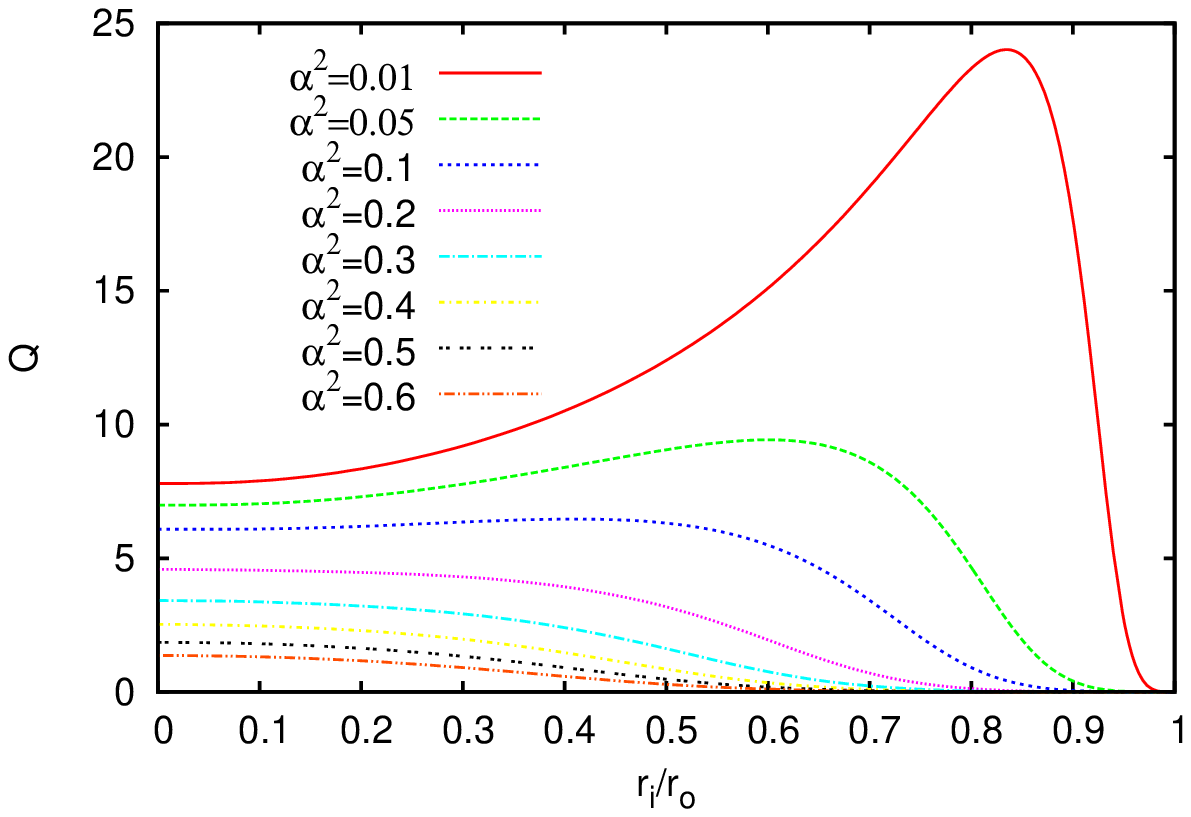}
\label{dShb}
}
}
\end{center}
\vspace{-0.5cm}
\caption{Properties of gravitating boson shells for several values of the coupling constant $\alpha^2$:  
(a) shows plot of the ratio $r_{\rm i}/r_{\rm o}$ of the shell versus $\rm log_{10}(r_{o})$, 
(b) shows plot of the charge $Q$ of the shell versus $r_{\rm i}/r_{\rm o}$.
\label{dsh1}
}
\end{figure}

Also, it may be interesting to observe (cf. Fig. \ref{dSha}) that as the inner shell radius $r_i$ is increased, the outer radius $r_o$ keeps increasing up to a certain value and then the trend reverses and $r_o$ starts decreasing when we further increase the inner shell radius $r_i$ . However, the value of the ratio $r_i/r_o$ keeps increasing towards its limiting value, i.e., to unity, with continuous increase in the value of $r_{i}$. This is seen in Fig.\ref{dSha}, where the ratio $r_i/r_o$ is shown versus the outer radius $r_o$. Fig. \ref{dShb} shows the plot of charge $Q$ of the shell versus $r_{\rm i}/r_{\rm o}$ for several values of the coupling constant $\alpha^2$.

In summary, we have studied boson stars and boson shells 
in a theory of a complex scalar field 
with a particular self-interaction potential as defined in Eq.~(1),
coupled to a $U(1)$ gauge field and to Einstein gravity. 
Because of the choice of self-interaction,
this theory has a positive cosmological constant $(\Lambda = 4 \pi G m^2 a^2)$.

Localized self-gravitating solutions are found to come in two types, 
having either a ball-like or a shell-like charge density.  
In particular,
the scalar field is finite only in a compact ball-like 
or shell-like region, 
whereas outside these regions the scalar field is identically zero.

We have studied the properties of these solutions 
and have determined their domains of existence 
for a set of values of the parameters of the theory. 
Similar but asymptotically flat solutions have been found before 
by Kleihaus, Kunz, L\"ammerzahl and List \cite{1,2}
for a V-shaped scalar potential. 

Here we have shown for the first time, that these charged shell-like solutions persist in the presence of a cosmological constant.
The self-gravitating compact boson shells constructed possess an empty de Sitter-like interior region, $r<r_i$, and a Reissner-Nordstr\"om-de Sitter exterior region, $r>r_o$. Our results represent a positive contribution in the direction of achieving the points mentioned in the motivations listed at the beginning of our work, justifying our present studies.

Further work could consider filling the interior region of the
boson shells with black holes, analogous to the study in \cite{2}.
Another interesting area to explore would be to extend these
charged compact boson shell solutions to other dimensions.

Towards the end we explain in brief, the numerical procedure used by us called as the \textit{shooting method}\cite{21}. In the shooting method, The given boundary value problem is transformed to initial value problem by choosing values for all dependent variables at one boundary. These values are arranged to depend
on arbitrary free parameters whose values are (initially) chosen in a random manner. We then
integrate the ODEs by initial value methods. Now we have a multidimensional root-finding problem. 

For example consider the BVP system (where $r_i<r<r_o $)
\[A'(r)=f_1(r,A(r),N(r),h(r),h'(r),b(r),b'(r)),~~~A(r_o)=1\]
\[b''(r)=f_2(r,A(r),N(r),h(r),h'(r),b(r),b'(r)),~~~b'(r_i)=0,~~~b'(r_0)=0\]  .
The shooting method looks for initial conditions $~A(r_i) = a_1$ and $b(r_i)=a_2$ so that $A(r_o) = 1$ $b'(r_i)=0$and $b'(r_o)=0$. Since we are varying the initial conditions, it makes sense to think of $A(r)$ and $b(r)$ as a function of $a_1$ and $a_2$, so shooting can be thought of as finding $a_1$ and $a_2$ such that: 
\[A'(r)=f(r,A(r),N(r),h(r),h'(r),b(r),b'(r)),~~~A(r_i)=a_1 ,~~~~A(r_o)=1\]
\[b''(r)=f_2(r,A(r),N(r),h(r),h'(r),b(r),b'(r))~~~b(r_i)=a_2,~~~b'(r_i)=0,~~b(r_o)=0\]
After setting up the function for $A$ and $b$, the problem is solved by the root finding methods to find the initial conditions $A(r_i)=a_1$ and $b(r_i)=a_2$ giving the roots.

We thank Jutta Kunz and Burkhard Kleihaus for introducing us to
this beautiful subject and for helpful discussions. We also thank the 
referee for his/her constructive comments and suggestions. 
One of us (SK) thanks the CSIR, New Delhi the award of a 
Junior Research Fellowship.


\begin{thebibliography}{50}
\bibitem{orbital}
C. Palenzuela, L. Lehner and S. L. Liebling,
Phys. Rev. D {\bf 77}, 044036 (2008)

\bibitem{gravity}
B. F. Schutz, Gravity from the Ground up, 3rd Ed (Cambridge Univ. Press) pp. 143 

\bibitem{Madsen}
 M.S. Madsen and A.R. Liddle,
Phys. Lett. B251 (4) 507

\bibitem{darkmatter}
R. Sharma, S. Karmakkar and S. Mukherjee,
[arXiv:0812.3470]

\bibitem{super1}

F. E. Schunck and A. R. Liddle, Lect. Notes Phys. {\bf 514}, 285 (1998) 

\bibitem{super2}
D.F. Torres, S. Capozziello and G. Lambiase,
Phys. Rev. D {\bf 62} (10), 104012 (2000) 

\bibitem{1} B. Kleihaus, J. Kunz, C. L\"ammerzahl and M. List, 
 Phys. Lett. {\bf B 675}, (2009) 102 
\bibitem{2} B. Kleihaus, J. Kunz, C. L\"ammerzahl and M. List, 
 Phys. Rev. {\bf D82},  (2010) 104050
\bibitem{3}
B.~Hartmann, B.~Kleihaus, J.~Kunz and I.~Schaffer
Phys.\ Rev.\ D {\bf 88}, 124033 (2013)

\bibitem{4}

B. Hartmann, B. Kleihaus, J. Kunz, I. Schaffer,
Phys. Let. {\bf B 714}, (2012) 120-126,

\bibitem{5}
D.~Astefanesei and E.~Radu,
Nucl.\ Phys.\ B {\bf 665} (2003) 594
[arXiv:gr-qc/0309131].

\bibitem{6} H. Arodz and J. Lis,
 Phys. Rev. {\bf D77}, (2008) 107702 [arXiv:0803:1566 [hep-th]].

\bibitem{7} B.~Kleihaus, J.~Kunz, M.~List and I.~Schaffer,   
Phys. Rev. D {\bf D77}, (2008) 064025 



\bibitem{9} P.~Jetzer,
Phys. Rept.  {\bf 220}, (1992) 163.

\bibitem{10} R.~Friedberg, T.~D.~Lee and A.~Sirlin,
Phys. Rev.  {\bf D13}, (1976) 2739.


\bibitem{11} S.~R.~Coleman, 
Nucl. Phys. {\bf B262}, (1985) 263
	


\bibitem{12} R. M. Wald, General Relativity (University of Chicago Press), Chicago (1984).
	
\bibitem{13} H.~Arodz, J.~Karkowski and Z.~Swierczynski,
 Phys. Rev. {\bf D80}, (2009) 067702

\bibitem{14}  H.~Arodz and J.~Lis, 
(arXiv: hep-th/0812.3284).

\bibitem{15} T.~D.~Lee and Y.~Pang, 
  Phys.\ Rept.\  {\bf 221}, (1992) 251.

\bibitem{16}
A.~Prikas,
Gen.\ Rel.\ Grav.\ {\bf 36} (2004) 1841
[arXiv:hep-th/0403019].



\bibitem{17} Jochum~J.~Van~Der~Bij,~Marcelo~Gleiser,
  Phys. Lett. {\bf B 194}, (1987) 482

\bibitem{18}
B.~Hartmann and J.~Riedel,
Phys.\ Rev.\ D {\bf 86}, 104008 (2012)

\bibitem{19}
E.~Radu and B.~Subagyo,
Phys.\ Lett.\ B {\bf 717}, 450 (2012)
[arXiv:1207.3715 [gr-qc]].


\bibitem{20}
B.~Hartmann and J.~Riedel,
Phys.\ Rev.\ D {\bf 87}, 044003 (2013)
[arXiv:1210.0096].


\bibitem{21}
http://reference.wolfram.com/mathematica/tutorial/NDSolveBVP.html

\end{thebibliography}
\end{document}